\newcount\draft\draft=0		
\newcount\cameraready\cameraready=1	

\ifnum\cameraready=1
 \documentclass[10pt]{sig-alternate-10pt}
\else
 \documentclass{sig-alternate}
\fi
\usepackage{graphicx,epsfig,sped,times,wide}
\usepackage{algorithm}
\usepackage[hyphens]{url}
\usepackage{clrscode}
\usepackage[noeka]{mathrmletter}
\usepackage{subfigure}
\usepackage{multirow}
\usepackage[normalem]{ulem}
\usepackage{soul}
\usepackage{multirow}
\usepackage{appendix}
\usepackage[noend]{algorithmic}
\usepackage{verbatim}
\usepackage{bm}
\PassOptionsToPackage{hyphens}{url}\usepackage[bookmarks,colorlinks=false, plainpages = false,breaklinks]{hyperref}
\usepackage{breakurl}

\ifnum\draft=0
\fi

\newcommand{\name} {CompAir}

\newenvironment{Itemize}%
{\begin{itemize}%
\setlength{\itemsep}{0pt}%
\setlength{\topsep}{0pt}%
\setlength{\partopsep}{0pt}%
\setlength{\parskip}{0pt}}%
{\end{itemize}}
{\begin{itemize}%
\setlength{\itemsep}{0pt}%
\setlength{\topsep}{0pt}%
\setlength{\partopsep}{0pt}%
\setlength{\parskip}{0pt}}%
{\end{itemize}}
\makeatletter
  \newcommand\figcaption{\def\@captype{figure}\caption}
  \newcommand\tabcaption{\def\@captype{table}\caption}
\makeatother

\newcommand{\xqed}{\nobreak \ifvmode \relax \else
      \ifdim\lastskip<1.5em \hskip-\lastskip
      \hskip1.5em plus0em minus0.5em \fi \nobreak
      \vrule height0.75em width0.5em depth0.25em\fi}

\newcommand{\xref}[1]{\S\ref{#1}}

\newcommand{\textred}[1]{\textcolor{red}{#1}}
\ifx\noeditingmarks\undefined
   \newcommand{\pgwrapper}[2]{\textred{#1: #2}}
\else
   \newcommand{\pgwrapper}[2]{}
\fi

\usepackage{xxx} 

\makeatletter

\global\def\section{\@startsection {section}{1}{\z@}%
                                   {2ex \@plus 1ex \@minus .1ex}%
                                   {1ex \@plus.2ex}%
                                   {\normalfont\bfseries\scshape\fontsize{11}{13}\selectfont}}
\global\def\subsection{\@startsection{subsection}{2}{\z@}%
                                     {2ex\@plus 1ex \@minus .1ex}%
                                     {1ex \@plus .2ex}%
                                     {\normalfont\bfseries\fontsize{10}{12}\selectfont}}
\global\def\subsubsection{\@startsection{subsubsection}{3}{\z@}%
                                     {2ex\@plus 1ex \@minus .1ex}%
                                     {1ex \@plus .2ex}%
                                     {\normalfont\itshape\fontsize{10}{12}\selectfont}}

\global\def\@maketitle{%
  \newpage
  \begin{center}%
  \let \footnote \thanks
  \null
    \vskip -.3em%
    {\LARGE\bf \@title \par}%
    \vskip 1em%
    {\large
      \lineskip .5em%
      \begin{tabular}[t]{c}%
        \@author
      \end{tabular}\par}%
    \vskip 1em%
    {\large \@date}%
  \end{center}%
  \par
  \vskip 2em}

\begin{document}

\title{Over-the-air Function Computation in Sensor Networks}

\newcommand{\supsym}[1]{\raisebox{6pt}{{\footnotesize #1}}}

\author{
\begin{tabular}{ccc}
Omid Abari & Hariharan Rahul & Dina Katabi\\
\multicolumn{3}{c}{Massachusetts Institute of Technology}\\
\multicolumn{3}{c}{\{abari, rahul, dina\}@csail.mit.edu}\\
\end{tabular}
}

\date{}
\maketitle

\newcommand*\wrapletters[1]{\wr@pletters#1\@nil}
\def\wr@pletters#1#2\@nil{#1\allowbreak\if&#2&\else\wr@pletters#2\@nil\fi}
\newcommand\ie{\textit{i.e.}}
\newcommand\newchanges[1]{{#1}}

\begin{sloppypar}

{\bf Abstract --}
Many sensor applications are interested in computing a function over
measurements (e.g., sum, average, max) as opposed to collecting all
sensor data.  Today, such data aggregation is done in a cluster-head.
Sensor nodes transmit their values sequentially to a cluster-head
node, which calculates the aggregation function and forwards it to the
base station.  In contrast, this paper explores the possibility of
computing a desired function over the air. We devise a solution that
enables sensors to transmit coherently over the wireless medium so
that the cluster-head directly receives the value of the desired
function. We present analysis and preliminary results that demonstrate
that such a design yield a large improvement in network throughput.


\section{Introduction}
\label{sec:intro}
The last decade has seen significant advances in sensor technologies
and protocols, which has led to interest in large sensor deployments
for monitoring various physical quantities such as temperature,
pressure, humidity,
\textit{etc.}~\cite{Sunhee2011SWATS,Estrin1999nextcentury}.
Deployments of over 1000 nodes are emerging quickly and expected to
become common~\cite{ISA_sensor}. They are used for environmental
monitoring, seismic sensing, factory automation, and process
control. As the density and scale of sensor networks grows, it
becomes increasingly important to come up with data transmission
schemes that use the medium efficiently.

In many sensor applications, there is interest only in summary
properties of the data --such as average, sum, maximum, minimum,
\textit{etc.} -- and there is no need for continuously collecting all
sensor
values~\cite{intanagonwiwat2003directed,madden2002tag,heinzelman2002application,mahimkar2004securedav}. For
example, the application may be interested in computing the average
temperature in an area, or checking that the maximum temperature does
not exceed a threshold.  For such applications, transmitting all data
to the base station is inefficient.  To address this issue, much past
work in sensor networks has advocated data
aggregation~\cite{intanagonwiwat2003directed,madden2002tag,heinzelman2002application,mahimkar2004securedav}. These
schemes typically aggregate the measurements locally at intermediate
nodes. For example, the network may be divided into clusters. One node
is elected as a cluster-head. The other nodes in the cluster transmit
their data sequentially to the cluster-head, which computes the
desired function and forwards it to the base station. While this
improves the overall network throughput, the sequential transmissions
of sensor data to the cluster-head still consumes significant wireless
bandwidth.


In this paper, we ask if it is possible to aggregate the data over the
air, directly delivering the desired function. For example, can
sensors intelligently modify the data that they transmit, such that
when they transmit jointly, the cluster-head simply receives the
result of the function?  Such an approach can yield a large reduction
in bandwidth consumption. Specifically, it ensures that bandwidth
needs do not increase with the size and density of the network; rather
they stay limited by the number of bits in the desired function.


We propose \name, a novel technique to aggregate data over the
air. \name\ leverages the fact that the wireless channel can combine
signals in both a linear and non-linear manner.  Specifically, when
signals are transmitted concurrently, the wireless channel naturally
produces a linear combination of their values weighted by the
channels' coefficients. This allows us to manipulate the transmissions
to produce a variety of linear functions such as sum, average,
variance, \textit{etc.} The wireless channel can also be used to
compute the non-linear ``OR'' function, by having the receiver observe
whether there is any power on the channel (beyond the typical noise
level).  We show that this property can be used as a primitive to
compute more complex non-linear functions such as minimum, maximum,
and median.

The paper further analyzes the robustness of the computed functions to
channel noise and presents preliminary results from a USRP testbed
that demonstrate that such a design can yield dramatic bandwidth
savings in large sensor networks.

\section{Related Work}\label{sec:related}
Related work falls in three categories.
\vskip 0.05in\noindent \textbf{(a) Aggregated Wireless Sensor Networks
  :} Most data aggregation work can be classified as:
tree-based
~\cite{madden2002tag,xu2001geography,lindsey2002data,intanagonwiwat2003directed},
cluster-based
~\cite{heinzelman2002application,yao2003query,zhou2004hierarchical,mahimkar2004securedav}
or multipath
~\cite{nath2004synopsis,manjhi2005tributaries,chen2006localized}. At a
high level, these schemes works as follows. Nodes transmit their
values sequentially to a cluster-head or parent using unicast
transmissions and a scheduling mechanism, say TDMA, to avoid
collisions. The parent/cluster-head computes the aggregated function
and then transmits this aggregate upstream. In contrast, in \name, all
sensors transmit their data concurrently, and the parent/cluster-head
simply receives a single value corresponding to the aggregate
function. This can improve spectrum efficiency and data latency.

\vskip 0.05in\noindent \textbf{(b) Collision Decoding:} Our work is
motivated by recent wireless trends advocating concurrent
transmissions and collision decoding. Systems like
ZigZag~\cite{zigzag}, Buzz~\cite{Buzz}, Caraoke~\cite{Caraoke}, and compressive
sensing~\cite{bajwa2006compressive,CS_sensor}, treat wireless collisions as a
code and decode the bits despite interference. Other systems like
MegaMIMO and AirSync~\cite{rahul2012megamimo,AirSync} exploit coherent
transmissions from multiple nodes to eliminate interference. Past
systems however are interested in decoding all of the original
bits. In contrast, \name\ notes that, in many cases, the objective is
to compute a function over the bits, and devises a mechanism for
computing such functions over the air.

\vskip 0.05in\noindent \textbf{(c) Information Theoretical Approaches
  to Distributed Function Computation:} There has been a lot of recent
interest in distributed function computation from the information
theory
community~\cite{giridhar2006,ying2007d,appuswamy2011,ramamoorthy2013,goldenbaum2013robust}. This
work either does not take advantage of the properties of the wireless
channel and simply focuses on minimizing communication cost over a
wired network like the Internet, or does not exploit the ability of
nodes to transmit jointly in a coherent manner. Further, this work is
theoretical.  In contrast, \name\ fundamentally leverages the ability
of wireless sensors to transmit coherently to compute a variety of
functions, and has been prototyped in a testbed of USRPs.


\section{Design Scope}\label{sec:scope}
\name\ is a wireless system that allows a collection of sensors to
transmit their data concurrently such that the receiver receives over
the medium a function of the sensors' data, such as sum, max, min,
etc.

Before delving into the details, we clarify the scope of this
particular paper and assumptions underlying \name.
\begin{Itemize}
\item \name\ is targeted towards {\it large and dense} sensor networks,
  which incur a high overhead from collecting individual sensor
  measurements from all the sensors, and can therefore obtain
  significant benefit from over-the-air aggregation of these
  measurements. 

\item We describe how \name\ works in the context of aggregating data
  from multiple sensors at a single cluster-head. The approach
  naturally extends to a multi-level hierarchy, including aggregating
  data from multiple cluster-heads.

\item In this paper, we assume that sensors can transmit their data
  coherently (\textit{i.e.}, synchronized in time and phase). Sensors
  can do so using recently developed synchronization techniques such
  as AirShare~\cite{AirShare}. AirShare is a simple low-overhead
  system that synchronizes nodes by transmitting the reference clock
  over the air, providing a tool for generic distributed PHY
  protocols. We refer the reader to~\cite{AirShare} for 
  further details regarding AirShare.
\end{Itemize}

The next two sections explain the basic idea underlying over-the-air
function computation. We start with linear functions, then extend the
design to non-linear functions.

\section{Computing Linear Functions}
\label{sec:linear}
It is well known that the wireless channel can be modeled as a linear
system. \name\ uses this property to compute linear functions
over sensor measurements. We will start by explaining how to compute
the sum of the sensors' measurements. We then extend \name\ to other
linear functions.

If multiple sensors transmit their signals simultaneously and
coherently, the cluster-head receives a linear combination of these
signals. Specifically, let sensor $i$, $1 \leq i \leq N$ ($N$ is the
number of sensors), transmit the signal $x_i$, and let the channel
from sensor $i$ to the cluster-head be $h_i$. Then, the received
signal at the cluster-head is $R = \sum_{i=1}^N h_ix_i$. (For
simplicity of exposition, we omit the noise terms from the equations
in this section. \xref{sec:robustness} describes in detail the effect
of noise, and how \name\ is robust against it.)

In order for the cluster-head to receive the true sum of the values
$x_i$, each sensor needs to compensate for its channel. In
particular, each sensor needs to transmit a value 
\begin{equation}
\label{eq:chancomp}
z_i = \frac{x_i}{h_i}
\end{equation}
such that:
\begin{equation}
R = \sum_{i=1}^N h_iz_i = \sum_{i=1}^N x_i.
\label{eq:sum}
\end{equation}

Naively, computing the channel from each sensor to the cluster-head
would require each sensor to individually send a training signal to
the cluster-head, and for the cluster-head to then transmit the
measured channel to the sensor. Such a system would incur high
overhead, and negate the benefits of using \name!

Instead, \name\ uses channel reciprocity to allow each sensor to
measure its channel to the cluster head with very low overhead (i.e.,
it uses the fact that the forward and reverse channels are always the
same up to a constant multiplier due to differences in hardware
between the transmit and receive chains~\cite{qualcomm80211n}.)  As
mentioned in~\xref{sec:scope}, \name\ synchronizes the oscillators on
the sensors and the cluster-head using distributed synchronization
mechanisms, particularly AirShare~\cite{AirShare}.  Once the
oscillators are synchronized, each \name\ sensor performs a one-time
calibration of the channels to and from the cluster-head to determine
a calibration factor $K_i = \frac{h_i(0)}{g_i(0)}$, where $h_i(0)$ is
the initial channel (at the time of calibration) from sensor $i$ to
the cluster-head as described earlier, and $g_i(0)$ is the
corresponding channel from the cluster-head to the sensor. Note that
this calibration factor depends only on the hardware on the nodes, and
needs to be repeated only infrequently to update the calibration
factor.

Each aggregated transmission is initiated by a request packet from the
cluster-head. Each sensor measures the channel $g_i(t)$ \textit{from}
the cluster-head using the request packet. It can then compute its
channel $h_i(t)$ \textit{to} the cluster-head as $h_i(t) = K_i
g_i(t)$. It then substitutes this computed channel in
Eq.~\ref{eq:chancomp} to determine the transmitted values $z_i$. Note
that each sensor needs to know only its own channel value, and hence
this computation can be done completely locally. Further, the ongoing
channel measurement overhead is small and constant independent of the
number of sensors.

\subsection{Computing Other Functions}

Now that we have described how \name\ can compute the sum of
transmitted values over the air, we can extend it to a variety of
other functions, which are either linear or can be reduced to linear
computations.

\vskip0.01in\noindent\textbf{Arithmetic Mean:} The AM can be computed
as the sum divided by the total number of sensors.

\vskip0.01in\noindent\textbf{Geometric Mean:} The GM is itself not
linear, but can be reduced to linear computations over the
air. Specifically, the logarithm of the GM is simply the AM of the
logarithms of the original values. Therefore, each node transmits the
$\log$ of its value, and the cluster-head computes the AM as above. It
can then compute the antilog to recover the actual geometric mean. The
same idea can be used to compute the product of observed values,
rather than the sum.

\vskip0.01in\noindent\textbf{Weighted average:} Each sensor simply sends $w_i x_i$ where
  $w_i$ is the weight associated with that sensor.

\vskip0.01in\noindent\textbf{Count (predicate):}  Suppose the system wants to count the
  number of sensors whose readings satisfy a certain value, say, if
  the temperature exceeds a threshold $T$. In such a case, the cluster
  head sends the corresponding predicate (``temperature > T'') in its
  request. Every sensor evaluates the predicate locally, and if it is
  true, it sends the value $x_i = 1$. The cluster head then simply
  receives the sum of these values. Note that this function can also
  be used to count the total number of sensors by simply setting the
  predicate to \textit{true}.

\vskip0.01in\noindent\textbf{Variance:} The variance can be computed
as $E(x^2) - E^2(x)$, where E(.) is the expectation/mean function. The
system can use \name\ to compute the mean of $x_i^2$ (each sensor
transmits $x_i^2$), as well as the mean of $x_i$. The cluster head can
use these values to determine the variance.

\vskip0.01in\noindent\textbf{Regression:} There are several
applications that involve the cluster head determining the
distribution of the sensor measurements. For instance, the cluster
head might desire to determine the best linear fit to the observed
measurements when plotted against the coordinates of the sensor. The
coefficients of such a fit can be measured simply by each sensor
transmitting the relevant terms of the linear regression. For
instance, if the best linear fit of the observed values can be written
in the form $y = \alpha + \beta x$, and each sensor's location and
observation value pair is $(x_i, y_i)$, then we can simply compute the
following values: $E(xy)$, $E(x)$, $E(y)$, $E(x^{2})$ over the
air. The cluster head can then compute the best fit as $\beta =
\frac{E(xy) - E(x)E(y)}{E(x^2)-E^2(x)}$ and $\alpha = E(y) - \beta
E(x)$.

\section{Going Beyond Linear Functions}
\label{sec:nonlinear}
In this section, we extend \name\ to compute some non-linear functions
such as maximum and minimum. The basic idea with these functions is to
leverage the fact that the wireless channel also effectively provides
an OR function. In particular, it can differentiate the case of when
power is being transmitted on the medium by one or more devices, from
the case of no power being transmitted by any device.


Let us start by explaining how we compute the maximum.  We can combine
the bit representation of the data with the channel's ``OR'' function
to compute the maximum value across the sensors as follows.  Computing
the maximum proceeds in rounds, from the high order bit to the low
order bit. In the first round, every node that has a 1 in the MSB
transmits, while nodes that do not have a 1 in that bit stay
silent. The cluster-head then detects whether power is received in
that round. If so, it determines that there is at least one node in
the network that has a 1 in the MSB, and sets the MSB of its computed
maximum to 1. In the next round, it then sends a request asking only
nodes that had a 1 in the MSB to transmit, if they have a 1 in the
second most significant bit. If no nodes transmit in this round, the
cluster head determines that the second most significant bit of the
maximum is 0. The cluster head then initiates the third round, and so
on, till it has computed the last bit. In this manner, the cluster
head can determine the maximum value across all sensors in the
network. Note that the number of rounds in this computation is
determined only by the bit resolution of the measured values,
typically 8-12. This is significantly less than the number of sensors
transmitting to the cluster-head in dense sensor networks. Further,
note that unlike in the case of linear functions, \name\ sensors need
not compensate for the amplitude of the channel to ensure correct
computation of the ``OR'' function.


Similar to the maximum, \name\ can also compute the minimum. It does
so by computing the maximum of the one's complement of the measured
values. The one's complement, $v$, of an n-bit value $x$, is
$2^n-x-1$. Thus, computing the minimum over the $x$ is the same as
computing the maximum over the $v$. Further, $v$ can be obtained from
$x$ simply by complementing all the bits in the representation of
$x$. \name\ can therefore compute the minimum across all sensor values
$x_i$ simply by computing the maximum over the bitwise complement of
$x_i$, and then computing the bitwise complement of the computed
result at the cluster head.

Going beyond minimum and maximum, \name\ can also compute percentiles,
say, the median, using a binary search across the data. \name\ first
computes the minimum and maximum as described above, as well as the
total number of nodes as described in~\xref{sec:linear}. It then
computes the function $count(min < value < (min+max)/2)$ over the air,
as described in~\xref{sec:linear}. If this count is less than half the
total number of nodes, it moves the interval to the right,
\textit{i.e.}, it computes the function $count((min+max)/2 < value <
3*(min+max)/4)$. If not, it moves the interval to the left,
\textit{i.e.}, it computes $count((min+max)/4 < value < (min+max)/2)$,
and so on, till it determines the boundary below which 50\% of the
data lies. Note that while we have described a simplified algorithm
above, \name\ can achieve significantly higher performance since it
computes the actual counts of sensors with values in each interval in
each round. With prior knowledge of the distribution of the data,
\name\ can therefore intelligently adapt the width of its search
intervals, instead of simply halving the interval each time. For
instance, consider the case when data is uniformly distributed between
the minimum and maximum values. If the function $count(min < value <
(min+max)/2)$ yields a value, say, $0.6 N$, then \name\ can simply
pick its next query to be $count(min < value <
(0.5/0.6)*(min+max)/2)$, as the median is likely to be very close to
this new value.

\section{Robustness of \name\ to Noise}
\label{sec:robustness}

Noise is a critical factor in any wireless system. Thus, in this
section we investigate how \name\ interacts with noisy channels, and
the impact of noise on throughput gain.

\subsection{Linear Functions}
We consider the performance of \name\ in the computation of linear
functions, \textit{e.g.}, sum.  In a noise-free scenario, sensors
transmit their scaled measurements concurrently, and the cluster-head
receives, in one shot, the sum as in Eq.~\ref{eq:sum}.  Such a
simplified model however is likely to yield an erroneous sum in
practice due to two types of noise:
\begin{Itemize}
\item
First, each received value incurs additive noise, which is a
combination of the receiver's thermal noise and quantization error.
As a result, the received sum after one transmission is likely
incorrect, at least in its least significant bits.  To increase
robustness to this noise, \name\ repeats the transmission of the sum
$M_1$ times and the cluster-head averages the received values.
\item
Second, since the sensor measurements in Eq.~\ref{eq:sum} have to be
compensated by the channel before transmission, there is a
contribution to noise from errors in the compensation factor. These
errors are due to errors in the estimation of the channel $\hat{h_i}$
from each sensor to the cluster-head. To address these errors, the
cluster-head broadcasts a request packet with $M_2$ channel estimation
samples, where the $M_2$ samples are used by each sensor for averaging
its channel estimate.
\end{Itemize}

Since noise terms are independent, averaging reduces noise power by a
factor equal to the number of averaged terms. However, how should the
system pick the values of $M_1$ and $M_2$? And, how do these
repetitions impact \name's throughput gain over a traditional system
that performs the aggregation at a cluster-head?

We consider a simple model with a group of $N$ sensors communicating
with a cluster-head. We compare \name\ with a traditional transmission
scheme that uses similar transmission power and bandwidth, and that
achieves the same data resolution for the computed function. We define
a sample as the value from an ADC operating at the Nyquist rate,
\textit{i.e.}, twice the wireless bandwidth of the signal.

Let the traditional scheme require a transmission of $M_D$ samples
from each sensor to the cluster-head. The total number of
transmissions required by the traditional scheme to get data from all
$N$ sensors, therefore, is $NM_D$.

In contrast, in \name, The cluster-head transmits a request packet
with $M_2$ channel estimation samples to the sensors.  Each sensor $i,
1 \leq i \leq N$, estimates its channel to the cluster-head,
$\hat{h_i}$, using these $M_2$ channel estimation samples. The sensors
then jointly transmit their data to the cluster-head, with each sensor
compensating by its channel estimate $\hat{h_i}$. They repeat this
joint transmission $M_1$ times. The total number of samples
transmitted by \name, \textit{i.e.}, its total overhead, is $M_A = M_1
+ M_2$.  We can therefore compute the throughput gain provided by
\name\ as $\alpha = \frac{NM_D}{M_1+M_2}$.

But how large should $M_1$ and $M_2$ be?  The values of $M_1$ and
$M_2$ are chosen based on the desired bit resolution, $b$, of the
function. Specifically, in order to get a resolution of $b$ bits, the
SNR of the sum has to be $6.02b+1.76$dB (see~\cite{waldenADC} for
derivation). Note that the SNR in computing the sum is different from
the SNR of the channel due to repetition and averaging. 
%
Specifically, the SNR in computing the sum depends on the averaging
parameters $M_1$ and $M_2$, which are chosen to reduce the estimation
noise and increase the SNR. Let us refer to the SNR of the sum as the
effective SNR, which we will compute below.

Let the average transmit power of each sensor be $P$, and the average
additive Gaussian noise at the receiver $\sigma$.  Then, the channel
SNR experienced by a traditional system is, by definition, $SNR =
\frac{P}{\sigma^2}$.

\begin{figure*}
\centering
\begin{tabular}{cc}
\epsfig{file=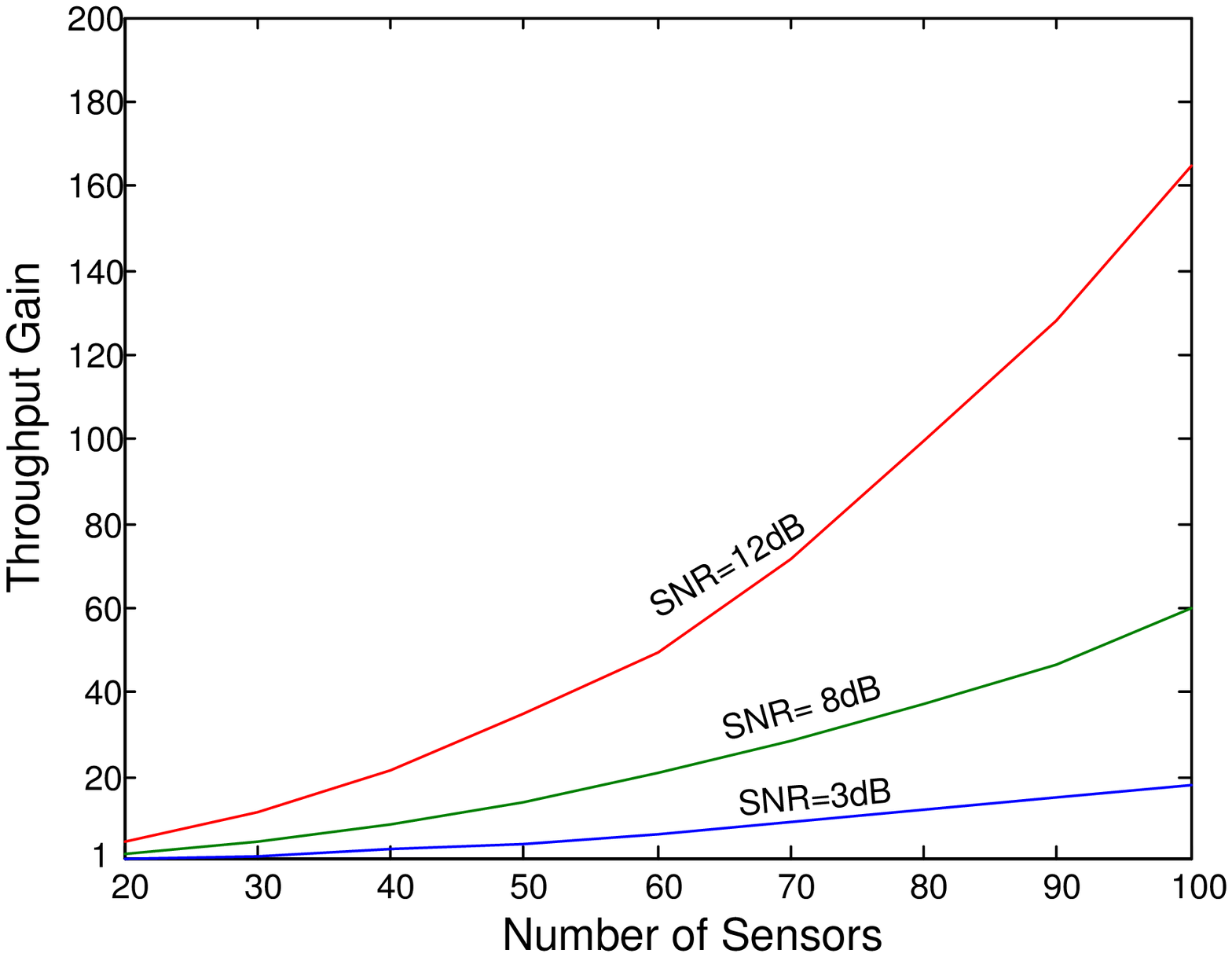,width=3in}&
\epsfig{file=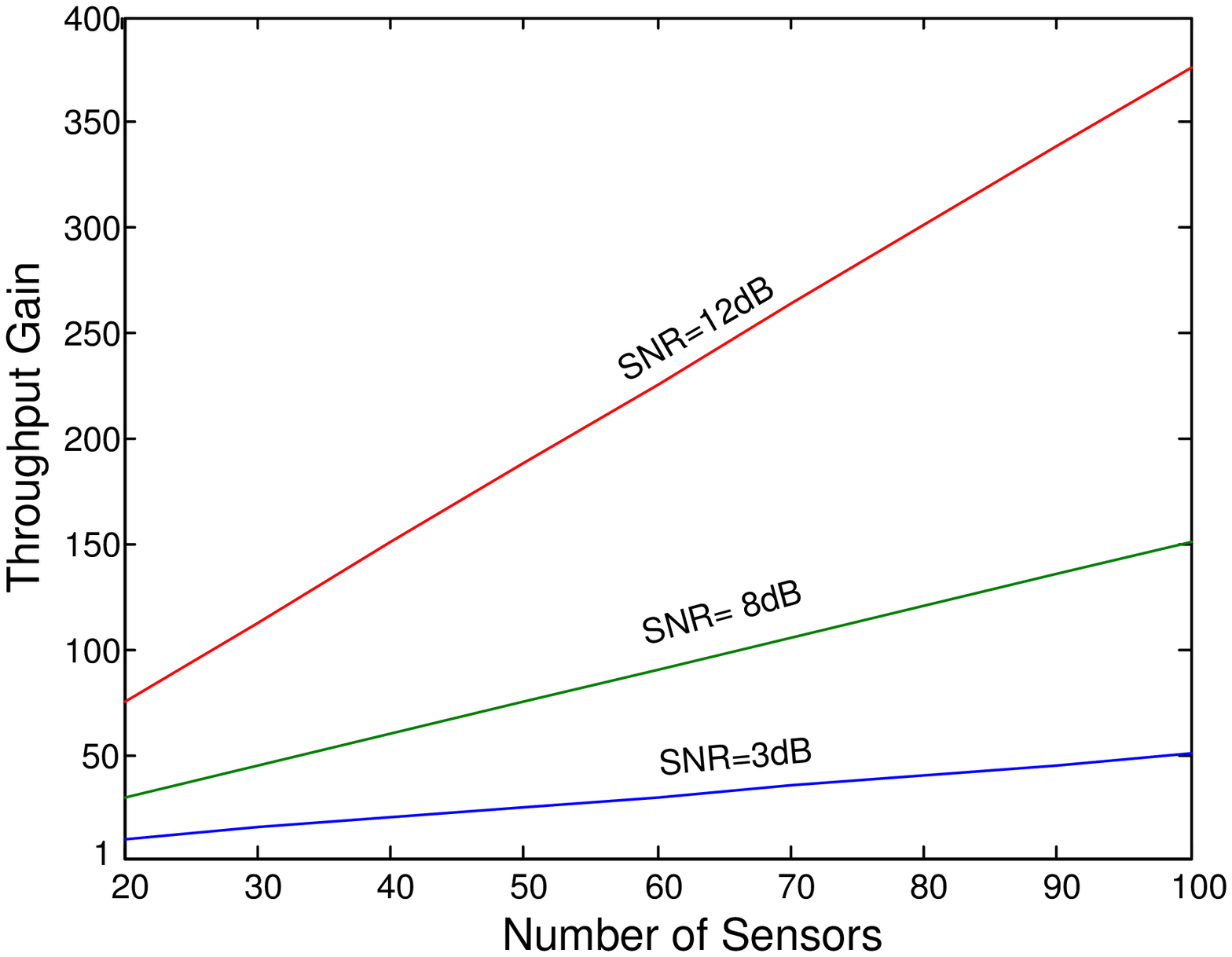,width=3in}\\
(a) Computing the $sum$ function &
(b) Computing the $\max$ function \\
\end{tabular}
\caption{\textbf{ \name\ Throughput Gain:} Figure shows that
  \name\ provides a throughput gain relative to Zigbee for computing
  functions across a wide range of channel SNR and different number of
  sensors.}
\label{fig:simgain}
\end{figure*}

In \name, however, sensors transmit jointly. Further their transmitted
samples add coherently since they are scaled by the magnitude and
phase of the channels. This translate into an SNR gain for the joint
transmission. Specifically, in each joint transmission, each sensor
transmits with an average power of $P$, \textit{i.e.}, the average
magnitude of the transmitted signal from each sensor is
$\sqrt{P}$. Since the sensors align their transmitted signals
coherently, the magnitude of the received signal at the cluster-head
is $N\sqrt{P}$, and consequently, the power of the received signal is
$N^2P$.\footnote{Note that, similar to beamforming, the fact that $N$
  signals, each with power $P$, produce a total power of $N^2P$ at the
  cluster-head does not violate conservation of power. This is
  because, the average power across all points in space still remains
  the same, and aligning the transmitted signals only reshapes the
  power profile to maximize the power at the cluster-head as if the
  $N$ sensors were an $N$-antenna MIMO transmitter that beamforms its
  signal to the cluster-head.}

When we average $M_1$ such joint transmissions, the average power
combines coherently whereas noise combines incoherently, because it is
independent across the different receptions. As a result, the average
received power after averaging $M_1$ joint transmission is $N^2P$ and
the average received noise power is $\frac{\sigma^2}{M_1}$.

There is a further noise contribution from errors in estimation of the
channel to the cluster-head. Since the channel estimate is averaged at
each sensor across $M_2$ samples, this reduces the expected noise in
the channel estimate at each sensor to
$\frac{\sigma^2}{M_2}$. Further, this noise adds up (incoherently as
before) across all the sensors in a joint transmission, producing a
total contribution of channel estimation noise equal to
$\frac{N\sigma^2}{M_2}$.

Combining these two, we therefore get that the total signal power is
$N^2P$ and the total noise power is $\frac{\sigma^2}{M_1} +
\frac{N\sigma^2}{M_2}$. The effective SNR of the system (i.e., the SNR
of the computed sum), therefore is the ratio of these two terms,
\textit{i.e.}, 
\begin{equation*}
SNR_{eff}  =  \frac{N^2P}{\frac{\sigma^2}{M_1} + \frac{N\sigma^2}{M_2}}
\end{equation*}
Since the original channel SNR is $\frac{P}{\sigma^2}$, we can
therefore write
\begin{equation*}
SNR_{eff}  = \frac{N^2 \times SNR}{\frac{1}{M_1} + \frac{N}{M_2}}
\end{equation*}
As described earlier, in order to get the desired $b$ bits of
resolution, the SNR requirement of the system can be written as
$dB(SNR_{eff}) = 6.02b+1.76$, where the function $dB(x) =
10\log_{10}(x)$. In order to maximize the throughput gain, therefore,
the \name\ cluster-head minimizes $M_1+M_2$ subject to the SNR
requirement. This convex optimization problem has a closed form
solution, and the cluster-head can therefore simply determine the
optimal values of $M_1$ and $M_2$.

Specifically, the optimal values are:
\begin{equation*}
 M_1 = \frac{1+\sqrt{N}}{lN}\text{ and } M_2 = \sqrt{N}M_1
\end{equation*}
where $l$ is defined by the equation $dB(l) = dB(N \times SNR) -
(6.02b+1.76)$, \textit{i.e.},
\begin{equation*}
l = N \times SNR \times 10^{\frac{-(6.02b+1.76)}{10}}
\end{equation*}
Based on these optimal values, we can compute the throughput gain of
\name\ as
\begin{equation*}
\alpha  =  \frac{N^3}{{(\sqrt{N}+1)}^2} \times SNR \times M_D \times 10^{\frac{-(6.02b+1.76)}{10}}
\end{equation*}

The throughput gain for linear-functions therefore scales linearly
with channel SNR (\textit{i.e.}, exponentially with SNR expressed in
dB), and approximately quadratically with the number of sensors.


\subsection{Non-linear Functions}

As described earlier, non-linear function computation proceeds in
rounds, with \name\ computing an ``OR'' function of the transmitted
bits in each round. The ``OR'' function is inherently robust because
it only requires differentiating the presence of power on the medium
from the absence of power (apart from receiver noise), and its
performance is a lower bound on any receivers which need to
detect power on the medium to begin decoding a packet. We defer the
analytical modeling of the performance of \name\ for non-linear
functions for a full version of the paper, and present simulation
results in~\ref{sec:eval}.

\section{Evaluation}
\label{sec:eval}
We present simulation results that show \name's scaling behavior for a
large number of sensors. We also present initial implementation
results using a USRP testbed to demonstrate that the design can be
built in real radios and its behavior matches the analysis.

\subsection{\name's Throughput Scaling}
\vskip0.05in\noindent{\textbf{Method:}} The chief promise of \name\ is
its ability to provide increasing throughput gains with increasing
density of sensors. In order to test \name's performance in the many
sensor regime, we build a sensor network simulation framework. We
deploy $N$ sensors in a $10m \times 10m$ grid (on the order of ZigBee
transmission range), with a single cluster-head for the grid.  We
increase the sensor density, by varying $N$ from 20 to 100. For each
$N$, we vary the average SNR from the sensors to the cluster-head. We
set the desired bit resolution to 8 bits and use the average SNR to
compute the parameters $M_1$ and $M_2$, as described
in~\xref{sec:robustness}. The receiver noise is generated using an
additive white Gaussian model, where the noise variance is scaled
according to the average SNR in the run.  We compare against a
traditional system that aggregates the measurements at the cluster-head
by having each sensor transmit its 8-bit measurement to the cluster
head using ZigBee. We do not simulate physical-layer headers and
ZigBee medium contention overhead. Both systems use 2~MHz bandwidth as
typical in ZigBee. The ZigBee system uses its typical transmission
rate of 250 kbps~\cite{Zigbee}. \name\ transmits in accordance with
the description in~\xref{sec:robustness} and~\xref{sec:linear} for
linear functions and the description in~\xref{sec:nonlinear} for
non-linear functions.  The ADC sampling rate for both systems is
assumed to be twice the bandwidth in accordance with the Nyquist
criterion.

\vskip0.05in\noindent{\textbf{Results:}} \textbf{Sum:}
Fig.~\ref{fig:simgain}(a) depicts the throughput gain of
\name\ relative to ZigBee for computing the $sum$ function while
varying the number of sensors and the average SNR. We see that
\name\ provides a throughput gain across the entire range of SNRs. The
throughput gain scales roughly quadratically with the number of
sensors. For example, for a dense network (\textit{i.e.}, 100 sensors
with an average of 12~dB), \name's throughput is $165\times$ that of
the traditional ZigBee system when computing a linear function. As
expected the gain is small for low density networks where the number
of sensors per cluster-head is about 20. Hence, \name\ is particularly
useful for dense large sensor networks.

\textbf{Max:} Fig.~\ref{fig:simgain}(b) plots the throughput gain for
computing the $\max$ function for a bit resolution of 8.  The
throughput gain scales linearly with the number of sensors. Note that
\name\ naturally provides a linear gain with the number of sensors
because the traditional system requires a linear increase in the
number of unicast transmissions as the number of sensors increases,
whereas the number of transmissions in \name\ stays fixed. Note that
linear functions provide an additional gain since all sensors
participate in all joint transmissions providing a corresponding
increase in power and hence SNR. In contrast, with $max$, only the
sensors with a bit value of 1 in the corresponding position
participate in a round, and the number of participating sensors
decreases as the number of rounds increases.


\subsection{Preliminary Implementation Results}
We show preliminary implementation results that demonstrate the
practicality of building \name\ in real radios.

\vskip0.05in\noindent{\textbf{Method:}} We implement a prototype of
\name\ using USRPs. The USRPs share a reference clock using
AirShare~\cite{AirShare}. The shared reference clock ensures that all
transmissions are coherent.  The implementation of AirShare follows
the description in~\cite{AirShare} and has been verified with the
authors of that paper.

We place the USRPs at random locations in an indoor testbed,
corresponding to locations of sensors and a cluster-head. The sensors
each uniformly pick random values between 0 and 1. We configure the
\name\ system to compute the $sum$ function across these values.
Specifically, the cluster-head initiates joint transmission with its
channel estimation packet. The sensors then compensate for the
estimated channel and transmit their scaled values coherently. The
cluster-head computes the received sum. The sensors keep repeating
their joint transmissions, and the cluster-head iteratively averages
the received values to update its computed sum. We compare this
computed sum from \name\ to the actual true value of the sum based on
the original value at each sensor.

Overall the experiment uses 6 USRPs as sensors and one USRP as a
cluster-head. However, since \name\ is particularly designed for a
large sensor network, we emulate the effect of many sensors by
repeating the experiment with a batch of 6 USRPs and combining their
received signals in post-processing. For example, to create a scenario
with 24 sensors we repeat the joint transmission of 6 sensors 4 times,
where each of these repetitions corresponds to different locations of
the USRPs. We then add the four received transmissions to create one
joint transmission of 24 sensors.  Note this is a conservative
estimate since we are adding up the noise added by receiver in four
experiments.

\begin{figure}
\centering
\begin{tabular}{cc}
\epsfig{file=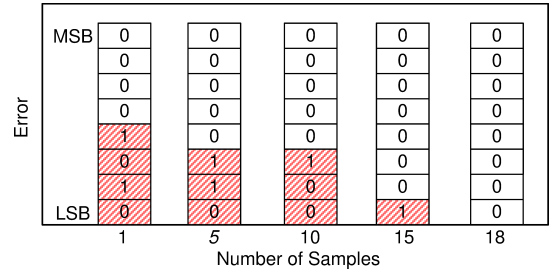,width=2.8in}\\ 
\end{tabular}
\caption{\textbf{Experimental Results for 24 Sensors:} This
  experiment shows that the noise affects the lower order bits of the
  sum value. The error goes down by averaging more samples. }
\label{fig:biterror}
\end{figure}

\vskip0.05in\noindent{\textbf{Result:}} Fig.~\ref{fig:biterror} shows
the results of computing an 8-bit sum over the measurements of 24
sensors.  The figure shows the the bit representation of the error
from MSB to LSB, where the error refers to the difference between the
sum computed at the cluster-head after running \name, and the actual
sum computed over the original sensor values.  As can be seen in the
figure, the errors are concentrated towards the LSB with the most
significant 4 of the 8 bits accurate even after 1 transmission. This
corresponds to our intuition that noise tends to affect the lower
order bits of the $sum$ value. Further, the figure shows that the
error goes down as we average more samples and eventually goes down
to 0. This shows that averaging reduces the impact of noise, and that
our \name\ implementation effectively delivers coherent transmission
across all sensors.

In this experiment, we need less than 18 samples to represent the sum
of value from 24 sensors with 8-bits of resolution. In this
experiment, the average SNR per sensor (\textit{i.e.}, the SNR when a
single sensor transmits alone) was about 3 to 4 dB.\footnote{The
  reason that the first transmission can deliver 4 correct bits is
  because the joint transmission has a higher power/SNR than
  individual transmission due to the sensors transmitting jointly and
  coherently.}  Thus, if each sensor were to transmit alone, the
system would have to use BPSK and each sensor out of the 24 sensors
would need at least 8 transmissions (ignoring coding) for a total of
192 transmissions. In contrast, \name\ computes the sum in less than
18 transmissions.

\vskip 0.5in
\section{Conclusion}
\label{sec:conclusion}

This paper proposes a novel over-the-air function computation
mechanism to compute a wide variety of popular linear and non-linear
functions over wireless sensor measurements. The mechanism leverages
simultaneous coherent transmission from multiple sensors and modifies
the sensor transmissions so that their collision on the wireless
channel produces the desired function value.

\let\oldthebibliography=\thebibliography
\let\endoldthebibliography=\endthebibliography
\renewenvironment{thebibliography}[1]{%
    \begin{oldthebibliography}{#1}%
      \setlength{\parskip}{0ex}%
      \setlength{\itemsep}{0ex}%
}%
{%
\end{oldthebibliography}%
}
{
\bibliographystyle{abbrv}
\bibliography{ourbib}

\begin{thebibliography}{10}

\bibitem{AirShare}
O.~Abari, H.~Rahul, D.~Katabi, and M.~Pant.
\newblock Airshare: Distributed coherent transmission made seamless.
\newblock In {\em IEEE INFOCOM}, 2015.

\bibitem{Caraoke}
O.~Abari, D.~Vasisht, D.~Katabi, and A.~Chandrakasan.
\newblock Caraoke: An e-toll transponder network for smart cities.
\newblock In {\em ACM SIGCOMM}, 2015.

\bibitem{appuswamy2011}
R.~Appuswamy and M.~Franceschetti.
\newblock Computing linear functions by linear coding over networks.
\newblock 2011.

\bibitem{bajwa2006compressive}
W.~Bajwa, J.~Haupt, A.~Sayeed, and R.~Nowak.
\newblock Compressive wireless sensing.
\newblock In {\em Proceedings of the 5th international conference on
  Information processing in sensor networks}, pages 134--142. ACM, 2006.

\bibitem{AirSync}
H.~V. Balan, R.~Rogalin, A.~Michaloliakos, K.~Psounis, and G.~Caire.
\newblock Airsync: Enabling distributed multiuser mimo with full spatial
  multiplexing.
\newblock 2012.

\bibitem{CS_sensor}
F.~Chen, F.~Lim, O.~Abari, A.~Chandrakasan, and V.~Stojanovic.
\newblock Energy-aware design of compressed sensing systems for wireless
  sensors under performance and reliability constraints.
\newblock {\em IEEE Transactions on Circuits and Systems I}, 2013.

\bibitem{chen2006localized}
S.~Chen and Z.~Zhang.
\newblock Localized algorithm for aggregate fairness in wireless sensor
  networks.
\newblock In {\em Proceedings of the 12th annual international conference on
  Mobile computing and networking}, pages 274--285. ACM, 2006.

\bibitem{Zigbee}
S.~C. Ergen.
\newblock {ZigBee/IEEE} 802.15.4 summary.
\newblock \url{http://pages.cs.wisc.edu/~suman/courses/838/papers/zigbee.pdf},
  2004.

\bibitem{Estrin1999nextcentury}
D.~Estrin, R.~Govindan, J.~Heidemann, and S.~Kumar.
\newblock Next century challenges: Scalable coordination in sensor networks.
\newblock In {\em Proceedings of the 5th Annual ACM/IEEE International
  Conference on Mobile Computing and Networking}, MobiCom '99, pages 263--270,
  New York, NY, USA, 1999. ACM.

\bibitem{giridhar2006}
A.~Giridhar and P.~Kumar.
\newblock Toward a theory of in-network computation in wireless sensor
  networks.
\newblock {\em Communications Magazine, IEEE}, 44(4):98--107, 2006.

\bibitem{goldenbaum2013robust}
M.~Goldenbaum and S.~Stanczak.
\newblock Robust analog function computation via wireless multiple-access
  channels.
\newblock {\em Communications, IEEE Transactions on}, 61(9):3863--3877,
  September 2013.

\bibitem{heinzelman2002application}
W.~B. Heinzelman, A.~P. Chandrakasan, and H.~Balakrishnan.
\newblock An application-specific protocol architecture for wireless
  microsensor networks.
\newblock {\em Wireless Communications, IEEE Transactions on}, 1(4):660--670,
  2002.

\bibitem{intanagonwiwat2003directed}
C.~Intanagonwiwat, R.~Govindan, D.~Estrin, J.~Heidemann, and F.~Silva.
\newblock Directed diffusion for wireless sensor networking.
\newblock {\em Networking, IEEE/ACM Transactions on}, 11(1):2--16, 2003.

\bibitem{ISA_sensor}
ISA.
\newblock {Industrial Wireless Sensor Networks: Trends and developments}.
\newblock \url{http://goo.gl/Rx5Vrl}, October 2012.

\bibitem{zigzag}
D.~Katabi and S.~Gollakota.
\newblock Zigzag decoding: Combating hidden terminals in wireless networks.
\newblock {\em ACM SIGCOMM Computer Communication Review}, 2008.

\bibitem{lindsey2002data}
S.~Lindsey, C.~Raghavendra, and K.~M. Sivalingam.
\newblock Data gathering algorithms in sensor networks using energy metrics.
\newblock {\em Parallel and Distributed Systems, IEEE Transactions on},
  13(9):924--935, 2002.

\bibitem{madden2002tag}
S.~Madden, M.~J. Franklin, J.~M. Hellerstein, and W.~Hong.
\newblock Tag: A tiny aggregation service for ad-hoc sensor networks.
\newblock {\em ACM SIGOPS Operating Systems Review}, 36(SI):131--146, 2002.

\bibitem{mahimkar2004securedav}
A.~Mahimkar and T.~S. Rappaport.
\newblock Securedav: A secure data aggregation and verification protocol for
  sensor networks.
\newblock In {\em Global Telecommunications Conference, 2004. GLOBECOM'04.
  IEEE}, volume~4, pages 2175--2179. IEEE, 2004.

\bibitem{manjhi2005tributaries}
A.~Manjhi, S.~Nath, and P.~B. Gibbons.
\newblock Tributaries and deltas: Efficient and robust aggregation in sensor
  network streams.
\newblock In {\em Proceedings of the 2005 ACM SIGMOD international conference
  on Management of data}, pages 287--298. ACM, 2005.

\bibitem{nath2004synopsis}
S.~Nath, P.~B. Gibbons, S.~Seshan, and Z.~R. Anderson.
\newblock Synopsis diffusion for robust aggregation in sensor networks.
\newblock In {\em Proceedings of the 2nd international conference on Embedded
  networked sensor systems}, pages 250--262. ACM, 2004.

\bibitem{rahul2012megamimo}
H.~Rahul, S.~Kumar, and D.~Katabi.
\newblock {MegaMIMO: Scaling Wireless Capacity with User Demands}.
\newblock In {\em ACM SIGCOMM 2012}, Helsinki, Finland, August 2012.

\bibitem{ramamoorthy2013}
A.~Ramamoorthy and M.~Langberg.
\newblock Communicating the sum of sources over a network.
\newblock {\em Selected Areas in Communications, IEEE Journal on},
  31(4):655--665, 2013.

\bibitem{qualcomm80211n}
System description and operating principles for high throughput enhancements to
  802.11.
\newblock IEEE 802.11-04/0870r, 2004.
\newblock {IEEE Standards Association}.

\bibitem{waldenADC}
R.~H. Walden.
\newblock Analog-to-digital converter survey and analysis.
\newblock {\em Selected Areas in Communications, IEEE Journal on},
  17(4):539--550, 1999.

\bibitem{Buzz}
J.~Wang, H.~Hassanieh, D.~Katabi, and P.~Indyk.
\newblock Efficient and reliable low-power backscatter networks.
\newblock {\em ACM SIGCOMM Computer Communication Review}, 42(4):61--72, 2012.

\bibitem{xu2001geography}
Y.~Xu, J.~Heidemann, and D.~Estrin.
\newblock Geography-informed energy conservation for ad hoc routing.
\newblock In {\em Proceedings of the 7th annual international conference on
  Mobile computing and networking}, pages 70--84. ACM, 2001.

\bibitem{yao2003query}
Y.~Yao and J.~Gehrke.
\newblock Query processing in sensor networks.
\newblock In {\em CIDR}, pages 233--244, 2003.

\bibitem{ying2007d}
L.~Ying, R.~Srikant, and G.~E. Dullerud.
\newblock Distributed symmetric function computation in noisy wireless sensor
  networks.
\newblock {\em Information Theory, IEEE Transactions on}, 53(12):4826--4833,
  2007.

\bibitem{Sunhee2011SWATS}
S.~Yoon, W.~Ye, J.~Heidemann, B.~Littlefield, and C.~Shahabi.
\newblock Swats: Wireless sensor networks for steamflood and waterflood
  pipeline monitoring.
\newblock {\em Network, IEEE}, 25(1):50--56, January 2011.

\bibitem{zhou2004hierarchical}
B.~Zhou, L.~H. Ngoh, B.~S. Lee, and C.~P. Fu.
\newblock A hierarchical scheme for data aggregation in sensor network.
\newblock In {\em Networks, 2004.(ICON 2004). Proceedings. 12th IEEE
  International Conference on}, volume~2, pages 525--529. IEEE, 2004.

\end{thebibliography}
}

\end{sloppypar}
\end{document}